\documentclass[10pt,conference]{IEEEtran}

\usepackage{lgrind}
\usepackage{epsfig}
\usepackage{amsmath}    
\usepackage{graphicx}
\usepackage{amssymb}
\pdfoutput=1

\begin{document}

\title{Classical Information Capacity of the Bosonic Broadcast Channel}

\author{
\authorblockN{Saikat Guha}
\authorblockA{Research Laboratory of Electronics \\
Massachusetts Institute of Technology \\
Cambridge, MA 02139 \\
saikat@MIT.edu}
\and
\authorblockN{Jeffrey H. Shapiro}
\authorblockA{Research Laboratory of Electronics \\
Massachusetts Institute of Technology \\
Cambridge, MA 02139 \\
jhs@MIT.edu}
}

\maketitle

\begin{abstract}
We show that when coherent-state encoding is employed in conjunction with coherent detection, the Bosonic broadcast channel is equivalent to a classical degraded Gaussian broadcast channel whose capacity region is dual to that of the classical Gaussian multiple-access channel. We further show that if a minimum output-entropy conjecture holds true, then the ultimate classical information capacity of the Bosonic broadcast channel can be achieved by a coherent-state encoding. We provide some evidence in support of the conjecture. 
\end{abstract}

\section{Introduction}

The past decade has seen several advances in evaluating classical information capacities of several important quantum communication channels \cite{bennettshor}--\cite{holevobook}. Despite the theoretical advances that have resulted \cite{bennettshor}, exact capacity results are not known for many important and practical quantum communication channels. Here we extend the line of research aimed at evaluating capacities of Bosonic communication channels, which began with the capacity derivation for the input photon-number constrained lossless Bosonic channel \cite{yuenozawa, caves}.   The capacity of the lossy Bosonic channel was found in \cite{ultcap},  where it was shown that a modulation scheme using classical light (coherent states) suffices to achieve ultimate communication rates over this channel. Subsequent attempts to evaluate the capacity of the noisy Bosonic channel with additive Gaussian noise \cite{holevobook} led to a crucial conjecture on the minimum output entropy of a class of Bosonic channels \cite{gglms}. Proving that conjecture would complete the capacity proof for the Bosonic channel with additive Gaussian noise, and it would show that this channel's capacity is achievable with classical-light modulation.  More recent work that addressed Bosonic multiple-access communication channels \cite{byen} revealed that modulation of information using non-classical states of light is necessary to achieve ultimate single-user rates.  In the present work, we study the classical information capacity of the Bosonic broadcast channel. A {\em {broadcast channel}} is the congregation of communication media connecting a single transmitter to two or more receivers. In general, the transmitter encodes and sends out independent information to each receiver in a way that each receiver can reliably decode its respective information. 

In Sec. II, we describe some recent work on the capacity region of the degraded quantum broadcast channel \cite{yard}. In Sec. III, we introduce the noiseless Bosonic broadcast channel model, and derive its capacity region subject to a new minimum output entropy conjecture. In Sec. IV we show that a recent duality result between capacity regions of classical multiple-input, multiple-output Gaussian multiple-access and broadcast channels \cite{goldsmith} does not hold for Bosonic channels.

\section{Quantum Degraded Broadcast Channel}

\begin{figure}
\begin{center}
\includegraphics[width=6.5cm,angle=0]{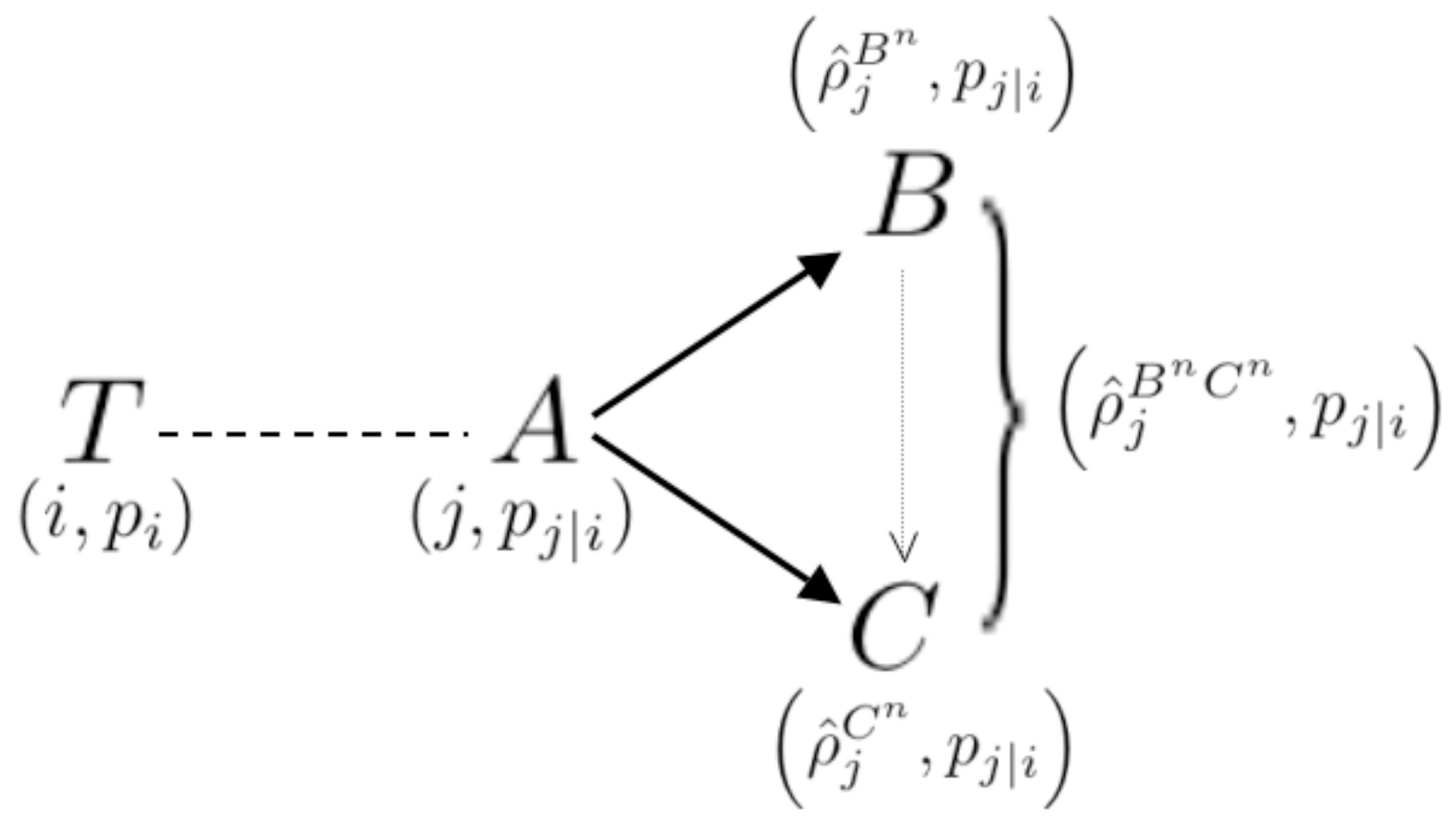}
\end{center}
\caption{Schematic diagram of the degraded single-mode Bosonic broadcast channel. The transmitter Alice ($A$) encodes her messages to Bob ($B$) and Charlie ($C$) in a classical index $j$, and over $n$ successive uses of the channel, prepares a bipartite state ${\hat \rho}_j^{B^nC^n}$ for them.}
\label{fig:TABC}
\end{figure}

A quantum channel ${\cal N}_{A-B}$ from Alice to Bob is a trace-preserving completely positive map that maps Alice's single-use density operators ${{\hat{\rho}}}^A$ to Bob's, ${{\hat{\rho}}}^B = {\cal N}_{A-B}({\hat{\rho}}^A)$. The two-user quantum broadcast channel ${\cal N}_{A-BC}$ is a quantum channel from sender Alice ($A$) to two independent receivers Bob ($B$) and Charlie ($C$). The quantum channel from Alice to Bob is obtained by tracing out $C$ from the channel map, i.e., ${\cal N}_{A-B} \equiv {\rm Tr}_C\left({\cal N}_{A-BC}\right)$, with a similar definition for ${\cal N}_{A-C}$. We say that a broadcast channel ${\cal N}_{A-BC}$ is {\em{degraded}} if there exists a {\em{degrading channel}} ${\cal N}^{\rm {deg}}_{B-C}$ from $B$ to $C$ satisfying
${\cal N}_{A-C} = {\cal N}^{\rm {deg}}_{B-C} \circ {\cal N}_{A-B}.$
The degraded broadcast channel describes a physical scenario in which for each successive $n$ uses of ${\cal N}_{A-BC}$ Alice communicates a randomly generated classical message $(m,k) \in (W_B, W_C)$ to Bob and Charlie, where the message-sets $W_B$ and $W_C$ are sets of classical indices of sizes $2^{nR_B}$ and $2^{nR_C}$ respectively. The messages $(m,k)$ are assumed to be uniformly distributed over $(W_B, W_C)$. Because of the degraded nature of the channel, Bob receives the entire message $(m,k)$ whereas Charlie only receives the index $k$. To convey these message $(m,k)$, Alice prepares $n$-channel use states that after transmission through the channel, result in bipartite conditional density matrices $\left\{{\hat \rho}_{m,k}^{B^nC^n}\right\}$, $ \forall (m,k) \in (W_B, W_C)$.  The quantum states received by Bob and Charlie, $\left\{{\hat \rho}_{m,k}^{B^n}\right\}$ and $\left\{{\hat \rho}_{m,k}^{C^n}\right\}$ respectively, can be found by tracing out the other receiver, viz.,  ${\hat \rho}_{m,k}^{B^n} \equiv {\rm Tr}_{C^n}\!\left({\hat \rho}_{m,k}^{B^nC^n}\right)$, etc. A $(2^{nR_B}, 2^{nR_C}, n, \epsilon)$ code for this channel consists of an encoder
\begin{equation}
x^n: (W_B,W_C) \rightarrow {\cal A}^n,
\label{eq:encoder}
\end{equation}
a positive operator-valued measure (POVM) $\left\{\Lambda_{mk}\right\}$ on ${\cal B}^n$ and a POVM $\left\{\Lambda_{k}^\prime\right\}$ on ${\cal C}^n$ which satisfy\footnote{${\cal A}^n$, ${\cal B}^n$, and ${\cal C}^n$ are the $n$ channel use alphabets of Alice, Bob, and Charlie, with respective sizes $|{\cal A}^n|$, $|{\cal B}^n|$, and $|{\cal C}^n|$.} 
\begin{equation}
{\rm Tr}\left(\hat{\rho}_{x^n(m,k)}(\Lambda_{mk} \otimes \Lambda_{k}^\prime)\right) \ge 1-\epsilon
\label{eq:dec_condition}
\end{equation}
for every $(m,k) \in (W_B,W_C)$.   A rate-pair $(R_B,R_C)$ is {\em achievable} if there exists a sequence of $(2^{nR_B},2^{nR_C},n,\epsilon_n)$ codes with $\epsilon_n \rightarrow 0$. The classical {\em capacity region} of the broadcast channel is defined as the convex hull of the closure of all achievable rate pairs $(R_B, R_C)$. The classical capacity region of the two-user degraded quantum broadcast channel ${\cal N}_{A-BC}$ was recently derived by Yard et. al. \cite{yard}, and can be expressed in terms of the Holevo information \cite{holevo}, 
\begin{equation}
\chi\!\left({p_j, {  {\hat{\sigma}}}_j}\right) \equiv S\!\left(\sum_j{p_j{  {\hat{\sigma}}}_j}\right) - \sum_jp_jS({  {\hat{\sigma}}}_j),
\end{equation}
where $\left\{p_j\right\}$ is a probability distribution associated with the density operators ${  {\hat{\sigma}}}_j$, and $S({\hat{\rho}}) \equiv -{\rm {Tr}}({\hat{\rho}}\log{\hat{\rho}})$ is the von Neumann entropy of the quantum state ${\hat{\rho}}$. 
Because $\chi$ may not be additive, the rate region $(R_B, R_C)$ of the degraded broadcast channel must be computed by maximizing over successive uses of the channel, i.e., for $n$ uses
\begin{eqnarray}
R_B &\le& \sum_ip_i\chi\!\left(p_{j|i},{\cal N}_{A-B}^{\otimes n}({\hat \rho}_j^{A^n})\right)/n \nonumber \\
&=& \frac{1}{n}\left[\sum_ip_iS\!\left(\sum_jp_{j|i}{\hat \rho}_j^{B^n}\right) \right. \nonumber \\ 
&&\left. - \sum_{i,j}p_ip_{j|i}S\!\left({\hat \rho}_j^{B^n}\right)\right], \quad \text{and}  \label{eq:RB-bound_chi} \\
R_C &\le& \chi\!\left(p_i, \sum_jp_{j|i}{\cal N}_{A-C}^{\otimes n}({\hat \rho}_j^{A^n})\right)/n, \nonumber \\
&=& \frac{1}{n}\left[ S\!\left(\sum_{i,j}p_ip_{j|i}{\hat \rho}_j^{C^n}\right)\right. \nonumber \\
&&\left. - \sum_ip_iS\!\left(\sum_jp_{j|i}{\hat \rho}_j^{C^n}\right) \right], \label{eq:RC-bound_chi} 
\end{eqnarray}
where $j \equiv (m,k)$ is a collective index and the states $\left\{{\hat \rho}_j^{A^n}\right\}$ live in the Hilbert space ${\cal H}^{\otimes n}$ of $n$ successive uses of the broadcast channel. The probabilities $\left\{p_i\right\}$ form a distribution over an auxiliary classical alphabet ${\cal T}$, of size $|{\cal T}|$, satisfying $|{\cal T}| \le \min\left\{|{\cal A}^n|,|{\cal B}^n|^{2} + |{\cal C}^n|^{2} + 1\right\}$. The ultimate rate-region is computed by maximizing the region specified by Eqs.~\eqref{eq:RB-bound_chi} and \eqref{eq:RC-bound_chi}, over $\left\{p_i\right\}$, $\left\{p_{j|i}\right\}$, $\left\{{\hat \rho}_j^{A^n}\right\}$, and $n$, subject to the cardinality constraint on $|{\cal T}|$. Fig.~\ref{fig:TABC} illustrates the setup of the two-user degraded quantum channel.

\section{Noiseless Bosonic Broadcast Channel}

The two-user noiseless Bosonic broadcast channel ${\cal N}_{A-BC}$ consists of a collection of spatial and temporal Bosonic modes at the transmitter (Alice), that interact with a minimal-quantum-noise environment and split into two sets of spatio-temporal modes en route to two independent receivers (Bob and Charlie). The multi-mode two-user Bosonic broadcast channel ${\cal N}_{A-BC}$ is given by $\bigotimes_s{{\cal N}_{A_s-B_sC_s}}$, where ${{\cal N}_{A_s-B_sC_s}}$ is the broadcast-channel map for the $s$th mode, which can be obtained from the Heisenberg evolutions
\begin{eqnarray}
{\hat b}_s &=& {\sqrt {\eta_s}}\,{\hat a}_s + {\sqrt {1-\eta_s}}\,{\hat e}_s, {\quad} {\text {and}}\label{eq:BS-modebk} \\
{\hat c}_s &=& {\sqrt {1-\eta_s}}\,{\hat a}_s - {\sqrt {\eta_s}}\,{\hat e}_s, \label{eq:BS-modeck}
\end{eqnarray}
where $\{{\hat a}_s\}$ are Alice's modal annihilation operators, and $\{{\hat b}_s\}$, $\{{\hat c}_s\}$ are the corresponding modal annihilation operators for Bob and Charlie, respectively. The modal transmissivities $\{\eta_s\}$ satisfy $0 \le \eta_s \le 1$, $\forall s$, and the environment modes $\{{\hat e}_s\}$ are in their vacuum states. We will limit our treatment here to the single-mode Bosonic broadcast channel, as the capacity of the multi-mode channel can in principle be obtained by summing up capacities of all spatio-temporal modes and maximizing the sum capacity region subject to an overall input-power budget using Lagrange multipliers, cf.~\cite{holevobook}, where this was done for the capacity of the multi-mode single-user lossy Bosonic channel.

\begin{figure}
\begin{center}
\includegraphics[width=6cm,angle=0]{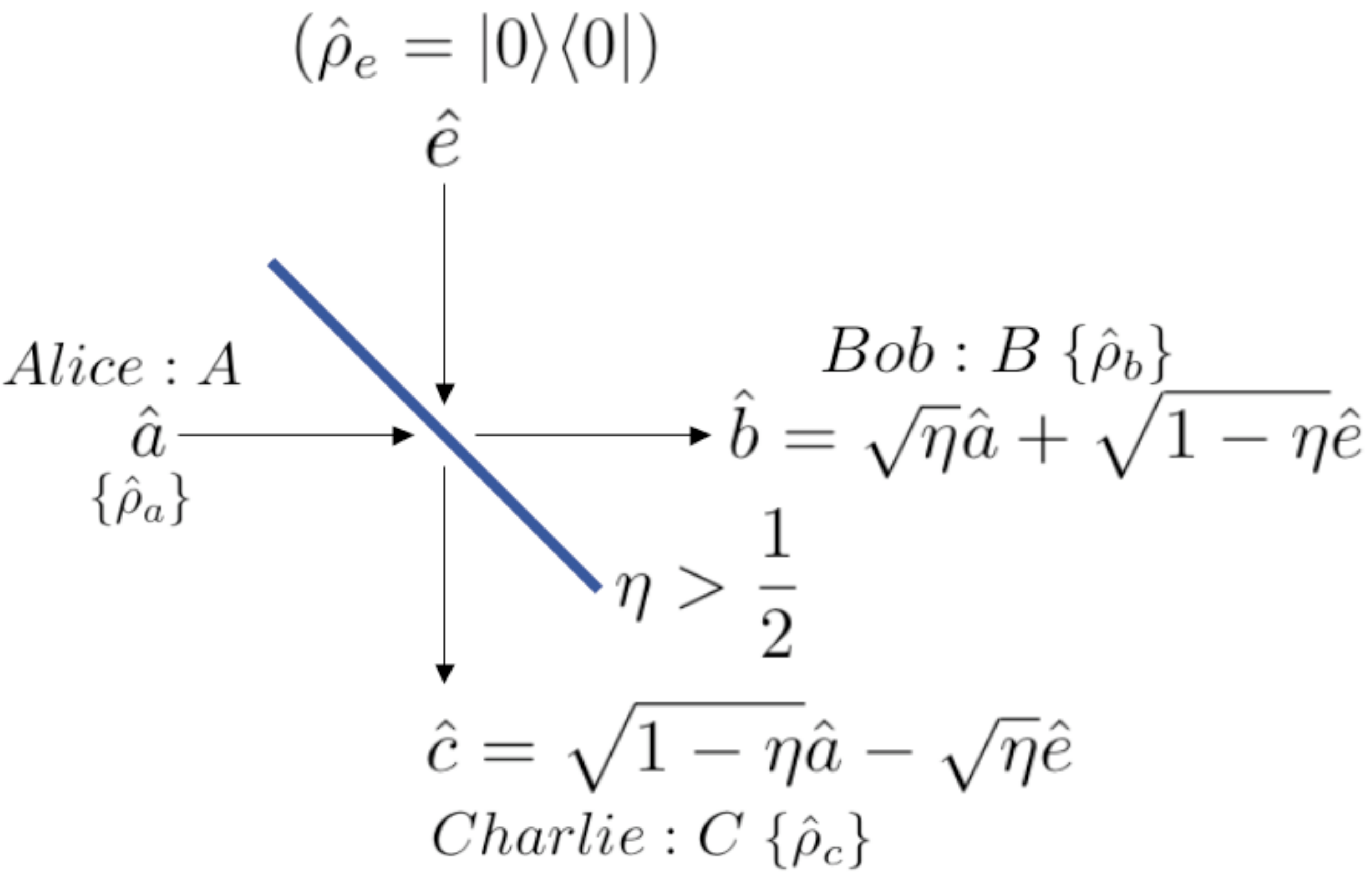}
\end{center}
\caption{A single-mode noiseless Bosonic broadcast channel can be envisioned as a beam splitter with transmissivity $\eta$. With $\eta > 1/2$, the Bosonic broadcast channel reduces to a `degraded' quantum broadcast channel, where Bob (B) is the less-noisy receiver and Charlie (C) is the more noisy (degraded) receiver.}
\label{fig:qbc}
\end{figure}

The principal result we have for the single-mode degraded Bosonic broadcast channel depends on a minimum output entropy conjecture (the strong form of Conjecture~2, see Appendix).  Assuming this conjecture to be true, we have that the ultimate capacity region of the single-mode noiseless Bosonic broadcast channel (see Fig.~\ref{fig:qbc}) with a mean input photon-number constraint $\langle{\hat a}^\dagger{\hat a}\rangle \le {\bar N}$ is
\begin{eqnarray}
R_B &\le& g(\eta\beta{\bar N}), {\quad} {\text {and}} \label{eq:RBultrate}\\
R_C &\le& g((1-\eta){\bar N}) - g((1-\eta)\beta{\bar N}), \label{eq:RCultrate}
\end{eqnarray}
for $0 \le \beta \le 1$, where $g(x) = (1+x)\log(1+x) - x\log(x)$.  This rate region is additive and achievable with single channel use coherent-state encoding with the distributions
\begin{eqnarray}
p_T(t) = \frac{1}{\pi{\bar N}}\exp\left(-\frac{|t|^2}{\bar N}\right), {\quad} {\text {and}}
\label{opt-dist1} \\  
p_{A|T}(\alpha|t) = \frac{1}{{\pi}{\bar N}\beta}\exp\left(-\frac{|{\sqrt {1-\beta}}\,t-\alpha|^2}{{\bar N}\beta}\right).
\label{opt-dist2}
\end{eqnarray}

\noindent {\bf Proof} --- It is straightforward to show that if $\eta > 1/2$, the Bosonic broadcast channel is a degraded quantum broadcast channel, in which Bob's is the less-noisy receiver and Charlie's is the more-noisy receiver. Yard et al.'s capacity region in Eqs.~\eqref{eq:RB-bound_chi} and \eqref{eq:RC-bound_chi} requires finite-dimensional Hilbert spaces. Nevertheless, we will use their result for the Bosonic broadcast channel, which has an infinite-dimensional state space, by extending it to infinite-dimensional state spaces through a limiting argument.\footnote{When $|{\cal T}|$ and $|{\cal A}|$ are finite, and we are using coherent states, there will be a finite number of possible transmitted states, which leads to a finite number of possible states received by Bob and Charlie.  Suppose we limit the auxiliary-input alphabet ($T$)---and hence the input ($A$) and the output alphabets ($B$ and $C$)---to truncated coherent states within the finite-dimensional Hilbert space spanned by the Fock states $\left\{|0\rangle, |1\rangle, \ldots, |K\rangle\right\}$, where $K \gg {\bar N}$. Applying Yard et al.'s theorem to the Hilbert space spanned by these truncated coherent states then gives us a broadcast channel capacity region that must be strictly an inner-bound of the rate-region given by unconditional equations~\eqref{coh-capacity1} and \eqref{coh-capacity2}. For made $K$ sufficiently large, while maintaining the cardinality condition, the rate-region expressions given by Yard et. al.'s theorem will converge to Eqs~\eqref{coh-capacity1} and \eqref{coh-capacity2}.} The $n$ = 1 rate-region for the Bosonic broadcast channel using a coherent-state encoding is thus:
\begin{eqnarray}
R_B &\le& \int{p_T(t)S\!\left(\int{p_{A|T}(\alpha|t)}|{\sqrt \eta}\,\alpha\rangle\langle{\sqrt \eta}\,\alpha|\,d\alpha\right)dt} \label{coh-capacity1} \\
R_C &\le& S\!\left(\int{p_T(t)p_{A|T}(\alpha|t)|{\sqrt {1-\eta}}\,\alpha\rangle\langle{\sqrt {1-\eta}}\,\alpha|\,d\alpha\,{dt}}\right) \nonumber \\ 
&-&{\int}p_T(t)S\!\left({\int}p_{A|T}({\alpha}|t)\right.  \nonumber \\
&\times& \left.|{\sqrt {1-\eta}}\,\alpha\rangle\langle{\sqrt {1-\eta}}\,{\alpha}|\,d\alpha\right)dt, \label{coh-capacity2}
\end{eqnarray}
where we need to maximize the bounds for $R_B$ and $R_C$ over all joint distributions $p_T(t)p_{A|T}(\alpha|t)$ subject to $\langle|\alpha|^2\rangle \le {\bar N}$. Note that $A$ and $T$ are complex-valued random variables, and the second term in the $R_B$ bound \eqref{eq:RB-bound_chi} vanishes, because the von Neumann entropy of a pure state is zero. Substituting Eqs.~\eqref{opt-dist1} and \eqref{opt-dist2} into Eqs.~\eqref{coh-capacity1} and \eqref{coh-capacity2}, shows that the rate-region Eqs.~\eqref{eq:RBultrate} and \eqref{eq:RCultrate} is achievable using single-use coherent state encoding.

For the converse, assume that the rate pair $(R_B, R_C)$ is achievable. Let $\left\{x^n(m,k)\right\}$, and POVMs $\left\{\Lambda_{mk}\right\}$ and $\left\{\Lambda_k^\prime\right\}$ comprise any $(2^{nR_B}, 2^{nR_C}, n, \epsilon)$ code in the achieving sequence. Suppose that Bob and Charlie store their decoded messages in the classical registers ${\hat W}_B$ and ${\hat W}_C$ respectively. Let us use $p_{W_B,W_C}(m,k) = p_{W_B}(m)p_{W_C}(k)$ to denote the joint probability mass function of the independent message registers $W_B$ and $W_C$.  As $(R_B,R_C)$ is an achievable rate-pair, there must exist $\epsilon_n^\prime \to 0$, such that
\begin{eqnarray}
nR_C &=& H(W_C) \nonumber \\
&\le& I(W_C; {\hat W}_C) + n\epsilon_n^\prime \nonumber \\
&\le& \chi(p_{W_C}(k),{\hat \rho}_k^{C^n}) + n\epsilon_n^\prime,
\end{eqnarray}
where $I(W_C; {\hat W}_C) \equiv H({\hat W}_C) - H({\hat W}_C|W_C)$ is the Shannon mutual information, and $\hat{\rho}^{C^n}_k = \sum_m p_{W_B}(m)\hat{\rho}_{m,k}^{C^n}$. The second line follows from Fano's inequality and the third line follows from Holevo's bound\footnote{Holevo's bound \cite{holevo}: Let $X$ be the input alphabet for a channel, $\left\{p_i, {\hat \rho}_i\right\}$ the priors and modulating states, $\left\{\Pi_j\right\}$ be a POVM, and $Y$ the resulting output (classical) alphabet. The Shannon mutual information $I(X;Y)$ is upper bounded by the Holevo information $\chi(p_i,{\hat \rho}_i)$}. Similarly, for an $\epsilon_n^{\prime\prime} \to 0$, we can bound $nR_B$ as
\begin{eqnarray}
nR_B &=& H(W_B) \nonumber \\
&\le& I(W_B;{\hat W}_B) + n\epsilon_n^{\prime\prime}  \nonumber \\
&\le& \chi(p_{W_B}(m),{\hat \rho}_m^{B^n}) + n\epsilon_n^{\prime\prime}  \nonumber \\
&\le& \sum_kp_{W_C}(k)\chi(p_{W_B}(m),{\hat \rho}_{m,k}^{B^n}) + n\epsilon_n^{\prime\prime},
\end{eqnarray}
where the three lines above follow from Fano's inequality, Holevo's bound and the concavity of Holevo information. In order to prove the converse, we now need to show that there exists a number $\beta \in [0,1]$, such that 
\begin{eqnarray}
\sum_kp_{W_C}(k)\chi(p_{W_B}(m),{\hat \rho}_{m,k}^{B^n}) \le ng(\eta\beta{\bar N}) \nonumber, \\
{\text{and}} {\quad}\chi(p_{W_C}(k),{\hat \rho}_k^{C^n}) \le ng((1-\eta){\bar N}) - ng((1-\eta)\beta{\bar N}). \nonumber
\end{eqnarray}
From the non-negativity of the von Neumann entropy $S\!\left({\hat \rho}_{m,k}^{B^n}\right)$, it follows that $\sum_kp_{W_C}(k)\chi(p_{W_B}(m),{\hat \rho}_{m,k}^{B^n}) \le\sum_kp_{W_C}(k)S\!\left(\sum_mp_{W_B}(m){\hat \rho}_{m,k}^{B^n}\right)$, as the second term of the Holevo information above is non-negative. Because the maximum von Neumann entropy of a single-mode Bosonic state with $\langle{\hat a}^\dagger{\hat a}\rangle \le {\bar N}$ is given by $g({\bar N})$, we have that
\begin{equation}
0 \le S\!\left({\hat \rho}_k^{B^n}\right) \le \sum_{j=1}^ng\!\left(\eta{\bar N}_{k_j}\right) \le ng\!\left(\eta{\bar N}_k\right),
\end{equation}
where, ${\bar N}_k \equiv \sum_{j=1}^n\frac1n{\bar N}_{k_j}$, and ${\bar N}_{k_j}$ is the mean photon number of the $j^{\rm {th}}$ symbol ${\hat \rho}_k^{B^n_j}$ of the $n$-symbol codeword ${\hat \rho}_k^{B^n}$, for $j \in \left\{1, \ldots, n\right\}$. Therefore, $\exists \beta_k \in [0,1]$, $\forall k \in W_C$, such that
\begin{equation}
S\left({\hat \rho}_k^{B^n}\right) = ng\left(\eta{\beta_k}{\bar N}_k\right).
\label{eq:eta-betak}
\end{equation}
Because of the degraded nature of the channel, Charlie's state can be obtained as the output of a beam splitter whose input states are Bob's state (coupling coefficient $\eta^\prime = (1-\eta)/{\eta}$ to Charlie) and a vacuum state (coupling coefficient $1-\eta'$ to Charlie). It follows, from assuming the truth of Strong conjecture~2 (see Appendix), that 
\begin{equation}
S\!\left({\hat \rho}_k^{C^n}\right) \ge ng\!\left((1-\eta){\beta_k}{\bar N}_k\right).
\label{eq:one-m-eta-betak}
\end{equation}
$\bar N$ is the average number of photons per-use at the transmitter (Alice) averaged over the entire codebook. Thus, the mean photon-number of the $n$-use average codeword at Bob, ${\hat \rho}^{B^n} \equiv \sum_kp_{W_C}(k){\hat \rho}_k^{B^n}$, is $\eta{\bar N}$. Hence, 
\begin{equation}
0 \le \sum_kp_{W_C}(k)S\!\left({\hat \rho}_k^{B^n}\right) \le S({\hat \rho}^{B^n}) \le ng\left(\eta{\bar N}\right),
\end{equation}
where the second inequality follows from the convexity of von Neumann entropy. The monotonicity of $g(x)$ then implies that there is a $\beta \in [0, 1]$, such that $\sum_kp_{W_C}(k)S\!\left({\hat \rho}_k^{B^n}\right) = ng(\eta\beta{\bar N})$. Hence we have,
\begin{equation}
\sum_kp_{W_C}(k)\chi(p_{W_B}(m),{\hat \rho}_{m,k}^{B^n}) \le ng(\eta\beta{\bar N}).
\label{eq:RBbeta}
\end{equation}
for some $\beta \in [0, 1]$. Equation~\eqref{eq:eta-betak}, and the uniform distribution $p_{W_C}(k) = 1/2^{nR_C}$ imply that
\begin{equation}
\sum_k\frac1{2^{nR_C}}g\left(\eta\beta_k{\bar N}_k\right) = g\left(\eta\beta{\bar N}\right).
\label{eq:thm1}
\end{equation}
Using \eqref{eq:thm1}, the convexity of $g(x)$, and $\eta > 1/2$, we have shown (proof omitted) that
\begin{equation}
\sum_k\frac1{2^{nR_C}}g\left((1-\eta)\beta_k{\bar N}_k\right) \ge g\left((1-\eta)\beta{\bar N}\right).
\label{eq:thm2}
\end{equation}
From Eq.~\eqref{eq:thm2}, and Eq.~\eqref{eq:one-m-eta-betak} summed over $k$, we then obtain
\begin{equation}
\sum_kp_{W_C}(k)S\left({\hat \rho}_k^{C^n}\right) \ge ng((1-\eta)\beta{\bar N}).
\label{eq:minoutent1}
\end{equation}
Finally, writing Charlie's Holevo information as
\begin{eqnarray}
\chi(p_{W_C}(k),{\hat \rho}_k^{C^n}) &=& S\!\left(\sum_kp_{W_C}(k){\hat \rho}_k^{C^n}\right) \nonumber \\
&&- \sum_kp_{W_C}(k)S\!\left({\hat \rho}_k^{C^n}\right) \nonumber \\
&\le& ng((1-\eta){\bar N}) \nonumber \\
&&- \sum_kp_{W_C}(k)S\!\left({\hat \rho}_k^{C^n}\right),  \label{eq:minoutent2}
\end{eqnarray}
we can use Eq.~\eqref{eq:minoutent1} to get
\begin{equation}
\chi(p_{W_C}(k),{\hat \rho}_k^{C^n}) \le ng((1-\eta){\bar N}) - ng((1-\eta)\beta{\bar N}),
\label{eq:RCbeta}
\end{equation}
which completes the proof.

\section{Discussion and Conclusion}

\begin{figure}
\begin{center}
\includegraphics[width=7cm,angle=0]{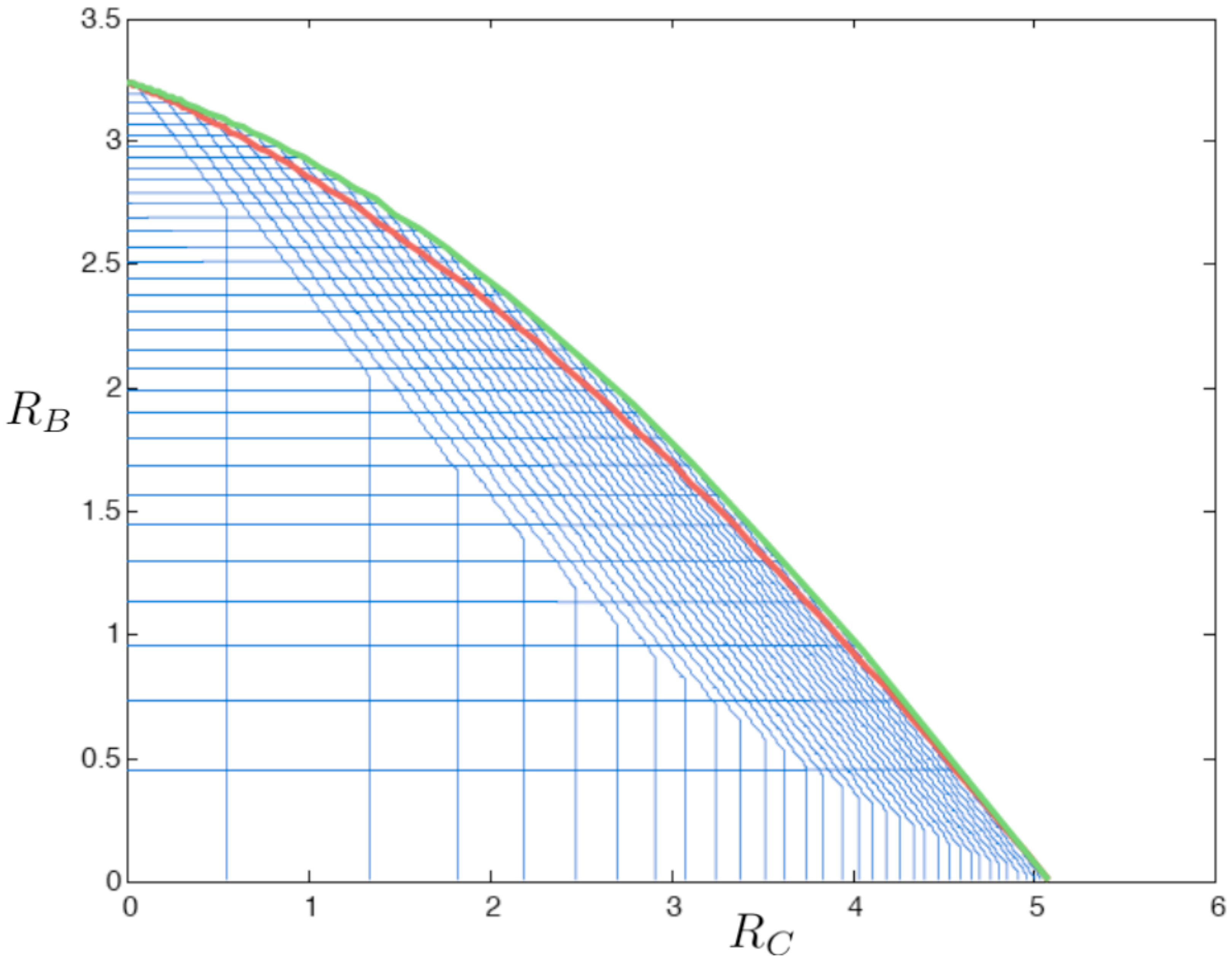}
\end{center}
\caption{Comparison of Bosonic broadcast and multiple-access channel capacity regions, in bits per channel use, for ${\eta = 0.8}$, and ${\bar N} = 15$.  The red line is the conjectured ultimate broadcast capacity region, which lies below the green line---the envelope of the MAC capacity regions.}
\label{fig:dualcheck}
\end{figure}

Recently, Vishwanath et. al. \cite{goldsmith} established a duality between the Òdirty paperÓ achievable region (recently proved to be the ultimate capacity region \cite{shamai2006}) for the classical multiple-input, multiple-output (MIMO) Gaussian broadcast channel and the capacity region of the MIMO Gaussian multiple-access channel (MAC).  The duality result states that if we evaluate the capacity regions of the MIMO Gaussian MAC---with fixed total received power $P$ and channel-gain values---over all possible power-allocations between the users, the corners of those capacity regions trace out the capacity region of the MIMO Gaussian broadcast channel with transmitter power $P$ and the same channel-gain values. Unlike this classical result, the capacity region of the Bosonic broadcast channel using coherent-state inputs is not equal to of the envelope of the MAC capacity regions using coherent-state inputs.  The capacity region of the Bosonic MAC using coherent-state inputs was first computed by Yen \cite{byen}.   In Fig.~\ref{fig:dualcheck} we compare the envelope of coherent-state MAC capacities to the capacity region of the coherent-state broadcast channel.   This figure shows that with a fixed beam splitter and identical average photon number budgets, more collective classical information can be sent when the beam splitter is used as a multiple-access channel as opposed to when it is used as a broadcast channel.

\begin{figure}
\begin{center}
\includegraphics[width=9cm,angle=0]{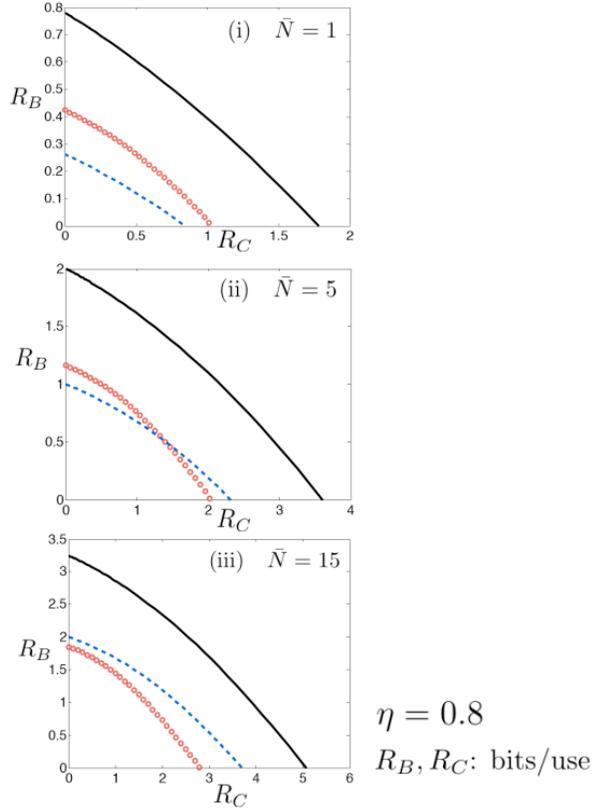}
\end{center}
\caption{Comparison of Bosonic broadcast channel capacity regions, in bits per channel use, achieved by coherent-state encoding with homodyne detection [{\em{red, circles}}], heterodyne detection [{\em{blue, dashed}}], and optimum reception [{\em{black, solid}}], for ${\eta = 0.8}$, and ${\bar N} = 1$, $5$, and $15$.}
\label{fig:capregionshomhet}
\end{figure}

The broadcast channel capacity region that we have derived---modulo Strong conjecture~2---exceeds what can be accomplished with conventional optical receivers, as shown in Fig.~\ref{fig:capregionshomhet}.  In this figure we compare the capacity regions attained by a coherent-state input alphabet using homodyne detection, heterodyne detection, and optimum reception. As is known for single-user Bosonic communications, homodyne detection performs better than heterodyne detection when the transmitters are starved for photons, because it has lower noise.  Conversely, heterodyne detection outperforms homodyne detection when the transmitters are photon rich, because it has a factor-of-two bandwidth advantage. To bridge the gap between the coherent-detection capacity regions and the ultimate capacity region, one must use joint detection over long codewords. Future investigation will need to be done to realize better broadcast communication rates over the Bosonic broadcast channel.

\section*{Acknowledgment}

This research was supported by the Defense Advanced Research Projects Agency. The authors thank Baris Erkmen for helpful discussions and for proving the Gaussian-state version of Strong conjecture~2.

\section*{Appendix: Minimum Output entropy Conjectures}

Let $\hat{a}$ and $\hat{b}$ denote the two input modes of a lossless beam splitter of transmissivity $\eta$, to produce output modes $\hat {c} = \sqrt{\eta} \,\hat{a} + \sqrt{1-\eta}\,\hat{b}$ and $\hat{d} = \sqrt{1-\eta}\,\hat{a} - \sqrt{\eta}\,\hat{b}$. In \cite{gglms}, we proposed the following minimum output entropy conjecture:

\noindent{\bf {Conjecture 1}} --- Let the input $\hat{b}$ be in a zero-mean thermal state with von Neumann entropy $S({\hat \rho}_B)=g(K)$. Then the von Neumann entropy of output $\hat{c}$ is minimized when $\hat{a}$ is in the vacuum state, and the minimum output entropy is $g((1-\eta)K)$.

In this paper, we propose a new output entropy conjecture:

\noindent {\bf {Conjecture 2}} --- Let the input $\hat{a}$ be in its vacuum state, input $\hat{b}$ in a zero-mean state with von Neumann entropy $S({\hat \rho}_B)=g(K)$. Then the von Neumann entropy of output $\hat{c}$ is minimized when $\hat{b}$ is in a thermal state with average photon number $K$, and the minimum output entropy is $g((1-\eta)K)$.

For the capacity proof of the Bosonic broadcast channel, we use Strong conjecture~2, which we now describe. Let the input modes $\{\hat{a}_i : 1\le i\le n\}$ be in a product state of $n$ vacuum states, and let the von Neumann entropy of the joint state of the inputs $\{\hat{b}_i: 1\le i\le n\}$  be $ng(K)$. Then, putting $\{\hat{b}_i : 1\le i\le n\}$ in a product state of mean-photon-number $K$ thermal states minimizes the output von Neumann entropy of the joint state of $\{\hat{c}_i: 1\le i\le n\}$.  Moreover, this minimum output entropy is $ng((1-\eta)K)$.

Previous work has provided considerable evidence in support of Conjecture~1 \cite{gglms}, \cite{renyiproof}.  In particular, we know that Conjecture~1 is true: when the state of $\hat{a}$ is  Gaussian;  when Wehrl entropy\footnote{The Wehrl entropy of a state with density operator $\hat{\rho}$ is the differential Shannon entropy of  $\langle \alpha|\hat{\rho}|\alpha\rangle/\pi$, where $|\alpha\rangle$ is a coherent state.} is considered instead of von Neumann entropy; and when R\'{e}nyi entropy of integer order $n\ge 2$ is considered instead of von Neumann entropy.  Strong conjecture~1, i.e., the $n$-use version, has been proven:  when the joint state of the $\{\hat{a}_i\}$ is Gaussian \cite{Eisert}; and when Wehrl entropy is considered instead of von Neumann entropy.  Other evidence in support of Conjecture~1 has been developed from entropy bounds \cite{gglms}, which show that the conjecture is asymptotically correct in the limit of weak and strong noise, and from simulated annealing starting with randomly selected initial states.  

In unpublished work, we have shown that Conjecture~2 is true:  when the state of $\hat{b}$ is Gaussian; when Wehrl entropy is considered instead of von Neumann entropy; and when the state of $\hat{b}$ is mixed and diagonal in the Fock basis with a probability distribution that is either Poisson, Binomial, or Bose-Einstein.  For Strong conjecture~2 we have shown that it is true:  when the $\{\hat{b}_i\}$ are in a Gaussian state; and when Wehrl entropy is considered instead of von Neumann entropy.


\begin{thebibliography}{99}

\bibitem{bennettshor} C. H. Bennett and P. W. Shor, ``Quantum information theory,'' {IEEE Trans. Inform.Theory} {\bf 44}, 2724--2742 (1998); A. S. Holevo, ``Coding theorems for quantum channels,'' { Tamagawa University Research Review} {\bf 4}, (1998), {quant-ph/9809023}; M. A. Nielsen and I. L. Chuang, {\em  Quantum Computation and Quantum Information} (Cambridge University Press, Cambridge, 2000).
\bibitem{yuenozawa} H. P. Yuen and M. Ozawa, ``Ultimate information carrying limit of quantum systems,'' {Phys. Rev. Lett.} {\bf 70}, 363--366 (1992).
\bibitem{caves} C. M. Caves and P. D. Drummond, ``Quantum limits on Bosonic communication rates," { Rev. Mod. Phys.} {\bf 66}, 481--537 (1994).
\bibitem{ultcap} V. Giovannetti, S. Guha, S. Lloyd, L. Maccone, J. H. Shapiro, and H. P. Yuen, ``Classical capacity of the lossy bosonic channel: the exact solution," {Phys. Rev. Lett.} {\bf 92}, 027902 (2004).
\bibitem{holevobook} V. Giovannetti, S. Guha, S. Lloyd, L. Maccone, J. H. Shapiro, B. J. Yen, and H. P. Yuen, ``Classical capacity of free-space optical communication,'' in O. Hirota, ed., {\em  Quantum Information, Statistics, Probability}, (Rinton Press, New Jersey, 2004) pp.~90--101.
\bibitem{gglms} V. Giovannetti, S. Guha, S. Lloyd, L. Maccone, and J. H. Shapiro, ``Minimum output entropy of bosonic channels: a conjecture," {Phys. Rev. A} {\bf 70}, 032315 (2004).
\bibitem{byen} B. J. Yen and J. H. Shapiro, ``Multiple-access bosonic communications," {Phys. Rev. A} {\bf 72}, 062312 (2005).
\bibitem{yard} J. Yard, P. Hayden, and I. Devetak, ``Quantum broadcast channels," {quant-ph/0603098}.
\bibitem{goldsmith} N. Jindal, S. Vishwanath, and A. Goldsmith, ``On the duality of Gaussian multiple-access and broadcast channels," { IEEE Trans. Inform. Theory} {\bf 50}, 768--783 (2004).
\bibitem{holevo} A. S. Holevo, ``The capacity of a quantum channel with general input states,'' {IEEE Trans. Inform. Theory} {\bf 44} 269--273 (1998); P. Hausladen, R. Jozsa, B. Schumacher, M. Westmoreland, and W. K. Wootters, ``Classical information capacity of a quantum channel,'' {Phys. Rev. A} {\bf 54}, 1869--1876 (1996); B. Schumacher and M. D. Westmoreland, ``Sending classical information via noisy quantum channels,'' {Phys. Rev. A}, {\bf 56}, 131--138 (1997).
\bibitem{shamai2006} H. Weingarten, Y. Steinberg, and S. S. Shamai, ``The Capacity region of the Gaussian multiple-input multiple-output broadcast channel," { IEEE Trans. Inform. Theory} {\bf {52}}, 3936--3964 (2006).
\bibitem{renyiproof} V. Giovannetti, S. Lloyd, L. Maccone, J. H. Shapiro, and B. J. Yen, ``Minimal R\'enyi and Wehrl entropies at the output of bosonic channels," {Phys. Rev. A} {\bf 70}, 022328 (2004).
\bibitem{Eisert}A. Serafini, J. Eisert, and M. M. Wolf, ``Multiplicativity and maximal output purities of Gaussian channels under Gaussian inputs,'' Phys. Rev. A {\bf 71}, 012320 (2005).


\end{thebibliography}
\end{document}